\documentclass[letterpaper]{article} 
\usepackage[draft]{aaai2026}  
\usepackage{times}  
\usepackage{helvet}  
\usepackage{courier}  
\usepackage[hyphens]{url}  
\usepackage{graphicx} 
\usepackage{amsfonts}
\usepackage{natbib}  
\usepackage{caption} 
\usepackage{algorithm}
\usepackage{algorithmic}
\usepackage{multirow}
\usepackage{amsmath}
\usepackage{booktabs}
\usepackage{array}
\usepackage{tikz}
\usepackage{newfloat}
\usepackage{listings}
\DeclareCaptionStyle{ruled}{labelfont=normalfont,labelsep=colon,strut=off} 
\floatstyle{ruled}
\newfloat{listing}{tb}{lst}{}
\floatname{listing}{Listing}

\usepackage{color}

\title{SurgPub-Video: A Comprehensive Surgical Video Dataset for \\ Enhanced Surgical Intelligence in Vision-Language Model}

\begin{document}
\author{Yaoqian Li\textsuperscript{1}\equalcontrib, Xikai Yang\textsuperscript{1}\equalcontrib, Dunyuan Xu\textsuperscript{1}\equalcontrib, Yang Yu\textsuperscript{1}, Litao Zhao\textsuperscript{1}, Xiaowei Hu\textsuperscript{2}, Jinpeng Li\textsuperscript{1}\thanks{Corresponding author: \texttt{jpli21@cse.cuhk.edu.hk}}, Pheng-Ann Heng\textsuperscript{1, 3}}
\affiliations{
    \textsuperscript{1}Department of Computer Science and Engineering, The Chinese University of Hong Kong, Hong Kong, China \\
    \textsuperscript{2}School of Future Technology, South China University of Technology, Guangzhou, China \\
    \textsuperscript{3}Institute of Medical Intelligence and XR, The Chinese University of Hong Kong, Hong Kong, China \\
   
}
\maketitle
\begin{abstract}
Vision-Language Models (VLMs) have shown significant potential in surgical scene analysis, yet existing models are limited by frame-level datasets and lack high-quality video data with procedural surgical knowledge. To address these challenges, we make the following contributions: (i) SurgPub-Video, a comprehensive dataset of over 3,000 surgical videos and 25 million annotated frames across 11 specialties, sourced from peer-reviewed clinical journals, (ii) SurgLLaVA-Video, a specialized VLM for surgical video understanding, built upon the TinyLLaVA-Video architecture that supports both video-level and frame-level inputs, and (iii) a video-level surgical Visual Question Answering (VQA) benchmark, covering diverse 11 surgical specialities, such as vascular, cardiology, and thoracic. Extensive experiments, conducted on the proposed benchmark and three additional surgical downstream tasks (action recognition, skill assessment, and triplet recognition), show that SurgLLaVA-Video significantly outperforms both general-purpose and surgical-specific VLMs with only three billion parameters. The dataset, model, and benchmark will be released to enable further advancements in surgical video understanding.

\end{abstract}

\section{Introduction}
\begin{figure}[!ht]
\centering
\includegraphics[width=\linewidth]{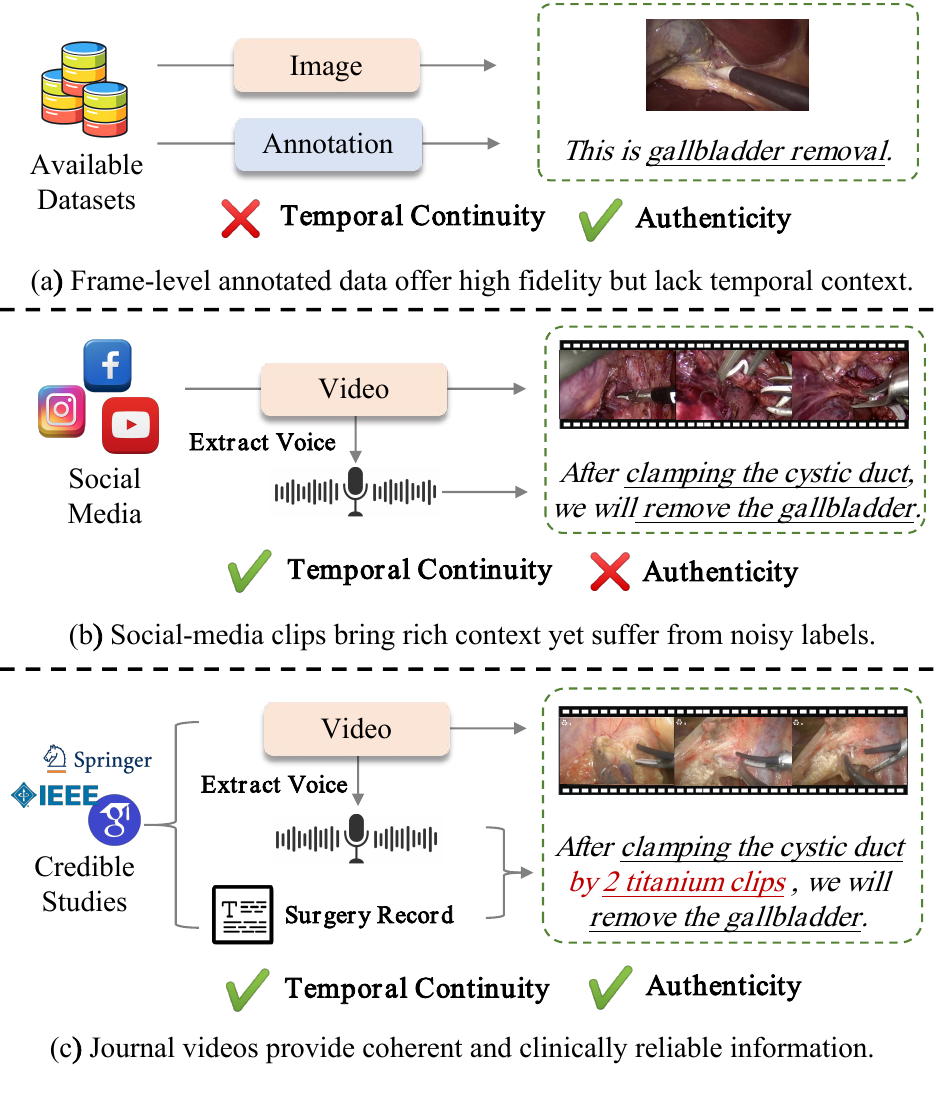}
\caption{Comparison of three pipelines for building surgical‑video VQA datasets. (a) Converting public frame‑label datasets offers high annotation fidelity but lacks narrative context. (b) Mining social media, e.g., YouTube, video narration supplies rich temporal descriptions yet suffers from inconsistent accuracy without peer review. (c) Leveraging peer‑reviewed journal videos combines continuous operative narration with authoritative reports, providing both temporal coherence and professional reliability.}
\label{teaser}
\end{figure}

Vision-Language Models (VLMs) have demonstrated exceptional performance across various domains~\cite{liu2023visualllava, li2023llavamed}. Recent work in surgical knowledge integration has shown promising results on tasks such as scene understanding, action recognition, and clinician skill assessment~\cite{zeng2025surgvlm,endobench}. 
However, current surgical VLMs are fundamentally constrained by their exclusive training on frame-level datasets~\cite{seenivasan2022surgical,yuan2024ssg}, thereby lacking the ability to understand temporal dynamics in surgical workflows.

\begin{table*}[h]
    \centering
    \addtolength{\tabcolsep}{0pt}
    \begin{tabular}{cccccccc}
    
        \toprule
        \multirow{2}{*}{Dataset}  &\multirow{2}{*}{Year}& \multirow{2}{*}{QA Level} & \multicolumn{2}{c}{Richness} & \multicolumn{2}{c}{Scale}&\multirow{2}{*}{Multi-Center}\\
\cmidrule(lr){4-5} \cmidrule(lr){6-7}        
        & & & \#Surgery Types &Language Source & \#Videos & \#Frames\\
        \midrule
        Cholec80-VQA &2022& Image & 1 & Categorical Text & 40 &21.6K&No \\
        EndoVis-18-VQA&2022  & Image & 1 & Categorical Text & 19 & 2K&No\\
        EndoVis-VQLA&2023 & Image & 1 & Categorical Text& 29 & 2.2K&No\\
        PSI-AVA-VQA&2024  & Image & 1 & Categorical Text&8 &2.2K&No \\
        SSG-VQA&2024  & Image & 1 & Categorical Text&50 & 25.5K&No\\
        Surg-396K &2024 & Image & 3 & Categorical Text & 109& 41.4K&Yes\\
        SurgVLM-DB&2025  & Image & 16 & Categorical Text& 849& 1.81M&Yes \\
        SurgPub-Video (Ours)&2025  & Video & 75 & Transcript \& Report  & 3,538 & 25M&Yes\\
        \bottomrule
    \end{tabular}
    \caption{Comparison of our SurgPub-Video with other surgical VQA datasets. Our SurgPub-Video is the only dataset specifically designed to support video-level surgical VQA, offering a unique advantage of temporal surgical scene understanding.}
    \label{tab:datasets}
\end{table*}


This limitation primarily stems from the scarcity of high-quality surgical video datasets with instruction-following annotations, hindering the advancement of surgical VLMs.
The curation of such datasets faces three key obstacles: 
(i) \textbf{Limited video diversity}: While frame-level surgical VQA datasets claim a large number of frames, they derive from a small pool of source videos, failing to capture the representative spectrum of surgical scenarios.
(ii) \textbf{Insufficient annotation granularity}: Surgical videos typically contain only sparse categorical labels, and converting them into detailed instruction-following samples requires extensive manual effort and is error-prone.
(iii) \textbf{Unreliable data sources}: Videos sourced from social media often lack credibility and accuracy, rendering them clinically unreliable.

Recent studies have explored a range of strategies to mitigate these challenges. 
To address the issue of data scarcity, some works attempt to collect existing public datasets and use frame-level annotations to construct surgical VQA datasets~\cite{zeng2025surgvlm,endobench}.
However, compared with native video-level labeling, converted annotations lack crucial temporal information, restricting the procedural understanding ability of surgical VLMs. 
Furthermore, only a limited number of surgical databases are entirely composed of video-based instances, 
and these often focus on pretraining image encoders using video captions~\cite{yuan2025learning, yuan2024procedure}. Consequently, these datasets do not include comprehensive question-answer pairs, which are essential for fine-tuning VLMs and improving their understanding of surgical tasks.
Another key challenge in current datasets is the lack of rich textual information for generating versatile QA pairs.
Since current image-level surgical datasets~\cite{seenivasan2022surgical, 10160403} usually provide only coarse category labels, such as “gallbladder removal” or “liver” for QA generation, omitting semantically rich descriptions of the visual scene. This limitation leads to VLMs aligning visual data with discrete labels, rather than with the open, descriptive space of natural language. For video-type data, a practical alternative is to leverage automatic speech recognition (ASR) models to generate transcripts as textual information~\cite{yuan2025learning}. Nevertheless, ASR outputs often contain noise and inconsistencies, requiring extensive post-processing.
Furthermore, the reliability of ASR transcripts is questionable, especially for videos not reviewed by clinicians, such as those sourced from social media platforms (e.g. YouTube)~\cite{yuan2024proceduresurgvlp,che2025surg3m}, where accuracy and authenticity are uncertain without expert validation.

To address these challenges, in this study, we construct the SurgPub-Video dataset, a high-quality surgical VQA dataset composed entirely of videos.
As shown in Fig.~\ref{teaser}, unlike previous surgical VQA datasets, our SurgPub-Video dataset exhibits superior qualities in terms of both temporal continuity and authenticity.
Specifically, we restrict our crawling scope to original surgical videos published in peer-reviewed articles, ensuring inherent vetting by the academic and clinical community for trustworthiness. As shown in Tab.~\ref{tab:datasets}, we collect a total of 3,538 surgical videos, significantly surpassing the size of previous surgical VQA datasets.
To create a comprehensive and high-quality surgical VQA database entirely composed of video clips, we move beyond categorical labels by utilizing both audio transcripts and associated surgical records to extract precise and semantically rich information.
We also design a sophisticated dataset curation workflow, including audio-guided video clip preprocessing, structured surgical concept extraction, and human-involved VQA pair creation. The final SurgPub-Video dataset contains 10,926 surgical clips and 48,520 VQA pairs, spanning 1,823 anatomical structures.
Moreover, considering current surgical VLMs heavily rely on the naive LLaVA structure, which only supports frame-level input and lacks explicit temporal modeling, to bridge this gap, we further introduce the SurgLLaVA-Video model, built upon the TinyLLaVA-Video architecture~\cite{zhang2025tinyllava}, which naturally supports whole-video input and leverages a resampler to integrate in-context relationships across the video. 
Finally, to comprehensively evaluate the capabilities of VLMs in surgical video understanding, we also propose the SurgPub-Video benchmark derived from our dataset, covering 11 vital surgical specialties and 5 main surgical tasks.

Our contribution can be summarized as follows:
\begin{itemize}
\item We introduce a high-quality, large-scale surgical dataset, SurgPub-Video, for VLM training, which contains VQA pairs with good temporal continuity, semantic richness, and authenticity.
\item We develop the SurgLLaVA-Video, which is trained on SurgPub-Video and enables both video-level and frame-level surgical scene understanding. 
\item We construct an extensive benchmark encompassing various surgical procedures and tasks. We systematically evaluate both open-source and commercial multimodal models on this benchmark and find that our SurgLLaVA-Video model achieves superior performance compared to existing dedicated and closed-source models.
\end{itemize}

\section{Related Work}

\begin{figure*}
\centering
\includegraphics[width=\linewidth]{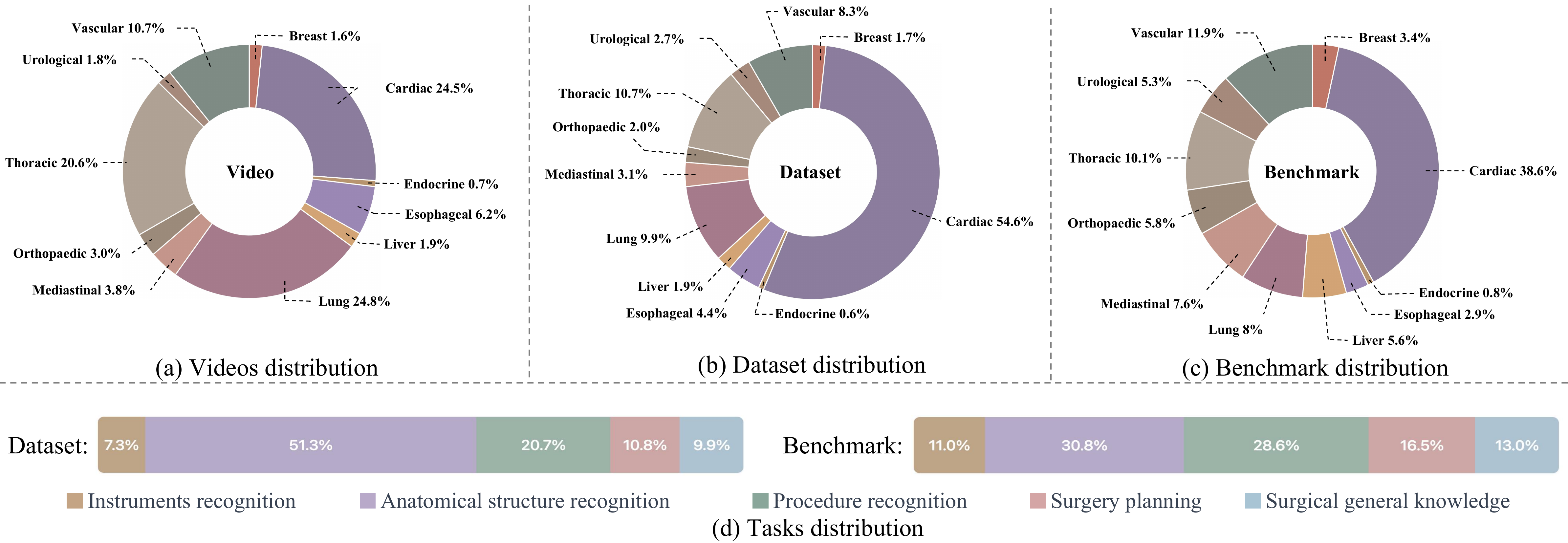}
\caption{Overview of the SurgPub-Video data distribution. (a) The proportion of original journal videos in 11 surgical specialties; (b) the specialty distribution of the SurgPub-Video; (c) The distribution of Benchmark specialties; (d) The coverage ratio of five types of VQA tasks (instrument, anatomy, procedure, planning, and general knowledge) in the Benchmark and complete datasets, demonstrating the credibility of data sources, the balance of specialties, and the diversity of tasks.}
\label{data_dis}
\end{figure*}
\subsection{Surgical Multimodal Datasets and Benchmarks}

A large-scale surgical VQA dataset is essential to enable vision language models to understand complex surgical scenarios.
However, traditional single-task image-level datasets with limited annotations struggle to support multiple organs and fine-grained recognition. 
Current VQA datasets are developed by converting single datasets, integrating multiple sources, or collecting network videos to address this gap. 
Early datasets, such as Cholec80-VQA~\cite{seenivasan2022surgical} and SSG-VQA~\cite{yuan2024ssg}, focus on converting single tasks like phase recognition to VQA pairs but lack sufficient annotations. 
To enhance diversity, SurgVLM-DB~\cite{zeng2025surgvlm} and EndoBench~\cite{endobench} aggregate multiple sources, covering various surgical procedures and organs. 
Recently, Surg-3M~\cite{che2025surg3m}, SurgVLP~\cite{yuan2024proceduresurgvlp}, and SurgVISTA~\cite{yang2025largesurgvista} collect network-sourced videos to create VQA datasets for multimodal tasks. 
Nevertheless, those manually annotated datasets~\cite{zeng2025surgvlm,endobench} cover limited procedures and organs, while network-sourced videos often lack peer-reviewed quality, and VQA annotations remain insufficient. 
These limitations restrict VLMs’ performance in complex surgical tasks. 
Our proposed work addresses these shortcomings by utilizing high-quality academic videos to construct a comprehensive VQA dataset.

\subsection{Surigcal Large Vision Language Model}
General VLMs, such as LLaVA~\cite{liu2023visualllava}, Qwen-VL~\cite{bai2025qwen2}, and CogVLM~\cite{wang2024cogvlm}, leverage instruction tuning on large-scale image-text pairs to excel in multimodal tasks like image captioning and visual question answering. 
Medical VLMs, including Med-PaLM~\cite{tu2024towardsmed-plam}, OmniMedVQA~\cite{hu2024omnimedvqa}, and LLaVA-Med~\cite{li2023llavamed}, utilize domain-specific datasets like MIMIC-CXR~\cite{johnson2019mimic} and SLAKE~\cite{liu2021slake} for pre-training and fine-tuning, enhancing performance in medical image analysis, diagnostic support, and VQA, particularly for radiology and pathology images. 
Surgical VLMs, such as SurgVLM~\cite{zeng2025surgvlm}, SurgicalGPT~\cite{seenivasan2022surgical}, and EndoChat~\cite{wang2025endochat} focus on processing individual surgical frames. 
These methods convert manually annotated datasets to VQA format and employ instruction tuning to support image-level tasks like phase recognition, action recognition, tool localization, and scene understanding.
%
%
However, due to the scarcity of large-scale, scenario-rich surgical video VQA datasets, current surgical VLMs remain restricted to static imagery and cannot perform video understanding across varied procedures. 
In this work, we address this gap by developing the SurgLLaVA-Video model, that trained on the SurgPub-Video dataset, and introducing video-level understanding to the VLM paradigm.
\section{Methodology}
In this section, we first present our constructed SurgPub-Video dataset and benchmark, followed by an introduction to the dataset curation workflow. Finally, we provide a description of our SurgLLaVA-Video model architecture.

\subsection{SurgPub-Video Dataset \& Benchmark}
\begin{figure*}[!ht]
\centering
\includegraphics[width=\linewidth]{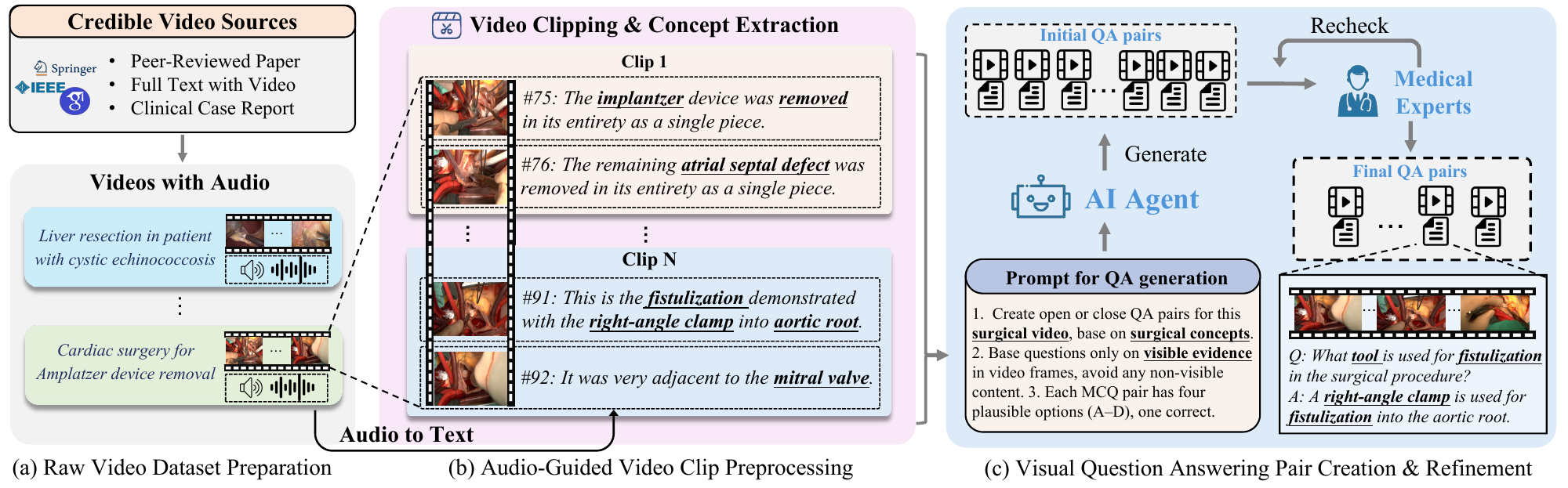}
\caption{VQA-generation pipeline for SurgLLaVA-Video dataset and benchmark. (a) Peer-reviewed journal videos and accompanying reports are collected. (b) Whisper transcripts and semantic filtering split each recording into coherent clips and extract core surgical concepts. (c) A prompt-based LLM agent produces initial open-ended and multiple-choice QA pairs for every clip; medical experts then review and refine these drafts, yielding the final high-quality VQA dataset.}
\label{framework}
\end{figure*}

\subsubsection{SurgPub-Video Dataset}
Tab.~\ref{tab:datasets} compares our constructed SurgPub-Video database against several popular surgical VQA databases.
Unlike most surgical databases that only provide single-frame snapshots, SurgPub-Video targets video-level analysis by consisting entirely of video clips, thereby preserving crucial temporal relationships that are essential for understanding surgical procedures. 
Fig.~\ref{data_dis} shows the distribution of the original video source, SurgPub-Video dataset, and the benchmark. Specifically, SurgPub‑Video consists of 10,926 surgical video clips with 29.85s average duration and 48,520 open and closed VQA pairs generated from five concept categories (instruments, procedural, anatomical structure, planning, general surgical knowledge). 
The clips span 11 surgical specialties and cover 75 kinds of surgery, 1,823 anatomical structures, 40 surgical instruments, and 1,290 unique procedures.


\subsubsection{SurgPub-Video Benchmark} We randomly sample 20\% of VQA pairs from 705 videos to construct the benchmark. To mitigate data imbalance and fairly evaluate the model's performance on different tasks, we removed some similar VQA pairs from the majority category. The final benchmark subset consists of 3,337 samples, reducing cardiac surgery samples to 38.6\%, and enhancing vascular and mediastinal surgeries to 11.9\% and 7.6\%, respectively. The benchmark assesses model performance through five VQA tasks, including instrument recognition, anatomical structure recognition, procedural step identification, surgical planning, and general surgical knowledge. The training set and test set are strictly video-level separated. Evaluation metrics include overall accuracy, specialty-specific accuracy, and task-specific accuracy, enabling detailed analysis of model capabilities and limitations across different surgical specialties and different tasks.


\subsection{Dataset Curation Workflow}
To construct a comprehensive and reliable surgical video-based VQA dataset, we first collect a substantial number of surgical videos from more than 25 peer-reviewed medical journals. 
After that, inspired by the VQA construction workflow of surgeons, we recruit and set a series of Large Language Models (LLMs) as Judgment Agents and design a systematic pipeline for generating video-based VQA pairs. Specifically, our approach consists of a coarse-to-fine video preprocessing strategy, beginning with coarse-grained video filtering based on audio transcripts (audio-to-text agent, coarse-grained video processing agent), then applying refined techniques to extract structured surgical concepts (fine-grained concept extraction agent). Finally, we employ a QA generation agent with strategically designed prompts to create VQA pairs. The detailed pipeline is described as follows.

\subsubsection{Raw Video Dataset Preparation}
To create a comprehensive and reliable surgical video database, all original videos in the SurgPub-Video dataset are crawled from 25 peer-reviewed medical journals. The complete list of journals we include is provided in the supplement. During our initial filtering process, we select articles accompanied by surgical videos and collect 3,538 raw surgical videos in total.
Fig.~\ref{data_dis} (a) presents the detailed composition of all included original videos. We collect videos from 11 surgery specialties. The average length of the video is 323.5 seconds. The original resolution of the video is 1920*1080, and the frame rate is 30Hz.
Compared to surgical videos from other sources such as YouTube, each video we collected is associated with a fully peer-reviewed article, ensuring its reliability for inclusion in our database. Moreover, each research article provides sufficient textual information related to its corresponding video.

\subsubsection{Audio-Guided Video Clip Preprocessing} 
Considering the significant redundancy within original surgical videos, we propose a coarse-to-fine video preprocessing pipeline that automatically identifies clips containing rich semantic information. We first introduce an audio-guided approach for coarse-level clip selection.
In particular, we employ an audio-to-text (ATT) agent powered by the OpenAI Whisper~\cite{radford2023robust} model to generate transcripts. While the ATT agent can segment the entire audio stream into short segments and convert them to timestamped text, the corresponding video fragments are often overly fragmented, many of which contain non-surgical content.
To address this limitation, we deploy a coarse-grained video processing (CGVP) agent to perform subsequent filtering and integration operations on these video fragments, ultimately producing video clips ranging from 15 to 30 seconds in duration. This agent removes redundant segments based on their transcribed content, such as non-surgical portions at the beginning and end of videos. Subsequently, the CGVP agent merges adjacent video fragments with similar or contextually related semantic content. This merging step is crucial because surgical procedures typically involve multiple sequential steps, such as vessel dissection (including fat clearance, hemostatic clamp placement, and vessel incision). The CGVP agent combines these related fragments to produce video clips that capture complete surgical procedures.
Fig.~\ref{framework} shows the detailed process of generating clips from raw video.

However, while the CGVP agent removes most of the redundancy from original videos, some video clips still contain unrelated information or unwanted surgical concepts. To address this issue, we additionally deploy a fine-grained concept extraction (FGCE) agent to determine whether video clips contain concepts of interest. We categorize surgical concepts into five types: instruments, procedural steps, anatomical structures, treatment planning, and general surgical knowledge. The FGCE agent prompts the LLM with the transcript associated with each video clip, along with textual information from the corresponding article, instructing the LLM to classify the video clip and organize the database into a structured format. This format records the length of each segment, text, and the concepts it contains.

\subsubsection{Visual Question Answering Pair Creation \& Refinement}
After completing the video clip cleaning process, QA generation (QAG) agent sequentially selects video clips and their corresponding textual information to construct QA pairs through targeted prompting. Specifically, to enhance clinical relevance, we design reasoning-based questions derived explicitly from causal explanations and rationales articulated within video narrations.
This approach ensures that generated questions possess authoritative clinical grounding, covering prediction of surgical outcomes, operative rationale, and risk-related decision-making processes.
For answer generation, AI agent produces both open-ended responses and multiple-choice questions with intentionally crafted distractors to reduce guessability.
All generated answers maintain medical accuracy, conciseness, and precise alignment with textual evidence, facilitating robust learning outcomes from the dataset.
We provide the abstract of prompt and an example of QA pair in Fig.~\ref{framework}.
We involve human expert reviews to refine the automatically generated VQA pairs, ensuring medical correctness and semantic consistency with the original intent. The iterative process between the QAG agent and human review may occur multiple times to progressively improve the quality of data samples within both the SurgPub-Video dataset and benchmark, ultimately establishing SurgPub-Video as an authoritative, clinically accurate resource for surgical VQA research.
\begin{figure}
\centering
\includegraphics[width=\linewidth]{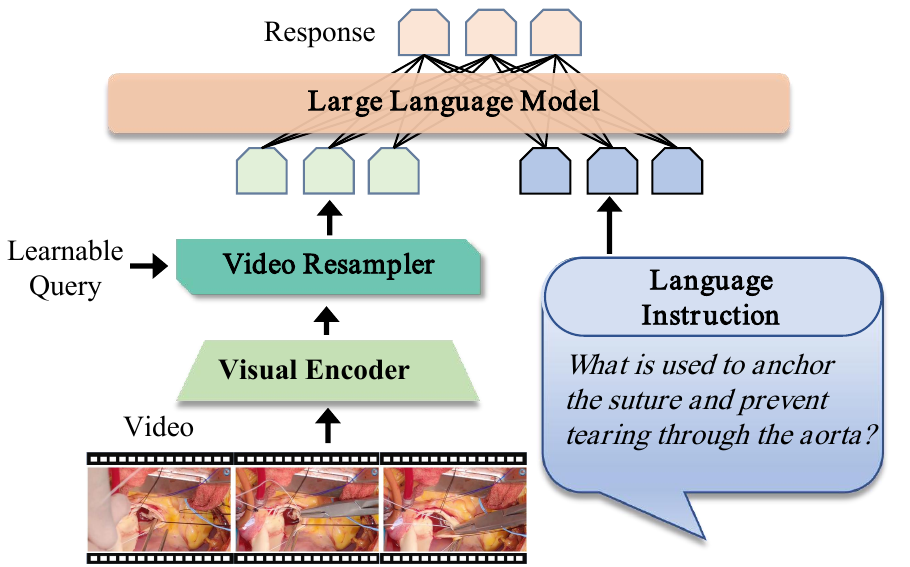}
\caption{Architecture of SurgLLaVA-Video. A multi-frame clip passes through a video encoder, whose patch tokens are compressed by a \textbf{video resampler}; learnable queries distill visual embeddings that are concatenated with the textual prompt and decoded by the LLM to generate the answer.}
\label{model}
\end{figure}
\subsection{Architecture of SurgLLaVA-Video Model}
Here, we present the architecture of our trained SurgLLaVA-Video model. 
Distinct from the recently proposed SurgVLM~\cite{zeng2025surgvlm}, which concentrates on frame-level analysis and directly concatenates visual tokens and text tokens as input to the LLM, SurgLLaVA-Video is particularly optimized for video-level input, which comprises three main components: a vision encoder, a video group resampler, and a LLM, following~\cite {zhang2025tinyllava}. Fig.~\ref{model} presents the model architecture, where the vision encoder first receives a video clip $X\in \mathbb{R}^{T \times H \times W \times 3}$ as input and extracts critical visual features at the frame level $Z\in \mathbb{R}^{T \times N \times D}$. $T$ denotes the video clip length, $H$ and $W$ represent the height and width of each frame, respectively, $N$ is the number of tokens corresponding to a single frame, and $D$ is the dimension of the extracted feature embedding. The extracted features $Z$ are then reshaped into $\hat{Z} \in \mathbb{R}^{(T \times N) \times D}$ for downstream operations.
Unlike the regular LLaVA structure, SurgLLaVA-Video incorporates a video resampler $\mathbf{P}_{\theta}$ that dynamically projects visual information into a fixed number of learnable queries $Q \in \mathbb{R}^{L \times D}$, simultaneously maintaining temporal relations across frames while saving computational resources. Inside the video resampler, due to the limited number of queries, we divide $\hat{Z}$ and $Q$ into multiple sub-embeddings $\{\hat{Z}_1, \hat{Z}_2, ... \hat{Z}_L\}$ and $\{Q_1, Q_2, ...Q_L\}$, respectively. Next, cross-attention operations are performed between each visual sub-embedding $\hat{Z}_i$ and corresponding learnable query $Q_i$, yielding $V_i$. We then concatenate $\{V_1, V_2, ...V_L\}$ to obtain the final visual input for the LLM: $V \in \mathbb{R}^{L \times D}$.
Finally, the language model combines the visual embeddings $V$ with user-provided surgical text questions $S$ as input and performs joint reasoning to generate responses.

When training the model, we use the pre-trained model of TinyLLaVA-Video and fine-tune it with the VQA pairs of SurgPub-Video. We only keep the visual encoder frozen, fully fine-tuning the LLM and video resampler.

\begin{table*}
\centering
\scriptsize
\renewcommand{\arraystretch}{0.8}

\fontsize{10pt}{10pt}\selectfont
\setlength{\tabcolsep}{1mm}
\begin{tabular}{cccccccccccccc}
\toprule
Model & Param. & Brst. & Card. & Endo. & Esop. & Liv. & Lung & Medi. & Orth. & Thor. & Urol. & Vasc. & Overall \\
\midrule
LLava-1.5 & 7B & 38.47&44.12&36.95 & 42.83&34.56&39.21 & 48.73&29.67&52.04 & 31.28&37.45&35.82\\
InternVL3 & 78B &75.00 & 78.72 & \underline{50.00} & 66.67 & 42.86 & 70.00 & 27.27 & 75.00 & 70.79 & 63.16 & 75.18 & 69.17 \\
Qwen2.5-VL-Instruct & 72B & \underline{81.82} & 61.24 & 38.46 & 60.00 & 33.33 & \underline{74.90} & 66.67 & 57.69 & 74.92 & 52.94 & 61.54 & 62.44 \\
GPT-4o-2024-0806 & - & \underline{81.82} & 67.98 & \underline{50.00} & 66.67 & 57.89 & 58.82 & 53.85 & 78.95 & 69.77 & \underline{77.78} & 62.00 & 66.67 \\
Qwen 2.5 Max & - & 75.00 & \underline{74.19} & 42.86 & 72.00 & \underline{74.03} & 71.53 & 68.18 & 65.66 & \textbf{86.39} & 67.65 & 73.58 & \underline{73.56 }\\
Gemini 2.0 Flash & - & 75.48 & 73.07 & 42.30 & 78.13 & 67.50 & 59.92 & \underline{72.67} & \textbf{81.46} & 78.30 & 75.93 & \underline{77.23} & 73.33 \\
SurgLLaVA-Video (Ours) & 3B & \textbf{82.14} & \textbf{84.38} & \textbf{61.38} & \textbf{89.69} & \textbf{74.19} & \textbf{82.84} & \textbf{79.76} & \underline{78.24} & \underline{81.74} & \textbf{80.23} & \textbf{82.73} & \textbf{82.84} \\

\bottomrule
\end{tabular}
\caption{Performance of various VLMs on proposed benchmark.}
\label{benchmark}
\end{table*}

\section{Experiments}

\begin{table*}[t]
\centering
\fontsize{10pt}{10pt}\selectfont
\renewcommand{\arraystretch}{1}
\setlength{\tabcolsep}{2mm}
\begin{tabular}{c|c|cccc|ccccc}
\toprule
\multirow{2}{*}{\raisebox{-1ex}{\centering Model}} & \multirow{2}{*}{\raisebox{-1ex}{\centering Param.}} & \multicolumn{4}{c}{RARP Action Recognition} & \multicolumn{4}{|c}{Endoscapes CVS Balanced Accuracy} \\
\cmidrule(lr){3-6} \cmidrule(lr){7-10}
& & Accuracy & Recall & Precision & Jaccard & Average & Crit. 1 & Crit. 2 & Crit. 3 \\
\midrule
LLava-1.5 & 7B & 23.46 & 14.22 & 11.53 & 6.31 & 49.56 & 48.97 & 50.00 & 49.72 \\
InternVL3 & 78B & 27.32 & 24.03& 30.48 & 13.30 & 53.50 & 52.02 & \underline{57.54} & \underline{50.94} \\
Qwen2.5-VL-Instruct & 72B & 28.20 & 13.04 & 22.30 & 5.94 & 47.22 & 50.26 & 41.67 & 49.72 \\
GPT-4o-2024-0806 & - & 28.10 & 15.99 & 16.15 & 9.16 & 9.07 & 9.23 & 8.81 & 9.17 \\
Qwen2.5-Max & - & 28.30 & 14.60 & 10.20 & 7.14 & 45.03 & 44.44 & 44.52 & 46.11 \\
Gemini 2.0 Flash & - & 24.40 & 17.51 & 18.54 & 7.27 & 52.37 & \underline{57.90} & 50.16 & 49.06 \\
SurgVLM & 72B & \underline{42.90} & \underline{34.64} & \underline{31.45} & \underline{19.22} & 51.40 & 51.59 & 52.22 & 50.39 \\
SurgLLaVA-Video (Ours) & 3B & \textbf{65.65} & \textbf{45.50} & \textbf{55.35} & \textbf{33.31} & \textbf{58.88} & \textbf{60.06} & \textbf{58.11} & \textbf{62.19} \\
\bottomrule
\end{tabular}
\caption{Performance of various VLMs on RARP action recognition and Endoscapes CVS task.}
\label{rarp&cvs}
\end{table*}

\begin{table*}[ht]
\centering
\fontsize{10pt}{10pt}\selectfont
\renewcommand{\arraystretch}{1}
\setlength{\tabcolsep}{3mm}
\begin{tabular}{c|c|cccc|ccccc}
\toprule
\multirow{2}{*}{\raisebox{-1ex}{\centering Model}} & \multirow{2}{*}{\raisebox{-1ex}{\centering Param.}} & \multicolumn{4}{c|}{Accuracy} & \multicolumn{4}{c}{mAP} \\
\cmidrule{3-6} \cmidrule{7-10}
& & Ins. & Verb & Target & Trip. & Ins. & Verb & Target & Trip. \\
\midrule
LLava-1.5 & 7B & 3.69 & 9.41 & 0.23 & 0.00 & 22.48 & 14.12 & 9.41 & 2.35 \\
InternVL3 & 78B & 38.14 & 9.52 & 3.29 & 0.52 & 22.74 & 14.08 & 9.44 & 2.36 \\
Qwen2.5-VL-Instruct & 72B & 32.66 & 7.91 & 5.25 & 1.27 & 23.38 & 14.45 & 10.86 & 2.71 \\
GPT-4o-2024-0806 & - & 13.33 & 5.89 & 5.94 & 1.50 & 22.80 & 14.34 & 10.42 & 2.60 \\
Qwen2.5-Max & - & 7.21 & 4.85 & 5.94 & 0.35 & 22.40 & 14.12 & 9.59 & 2.39 \\
Gemini 2.0 Flash & - & 15.18 & 7.44 & 15.87 & 1.85 & 23.14 & 14.45 & 10.91 & 2.54 \\
SurgVLM & 72B & \underline{62.20} & \underline{18.64} & \textbf{57.07} & \underline{12.52} & \underline{31.44} & \underline{19.16} & \textbf{21.01} & \underline{6.35} \\
SurgLLaVA-Video (ours) & 3B& \textbf{85.93} & \textbf{63.29} & \underline{47.12} & \textbf{39.30} & \textbf{58.98} & \textbf{42.75} & \underline{19.93} & \textbf{10.61} \\
\bottomrule
\end{tabular}
\caption{Performance of various VLMs on CholecT50 triplet recognition.  }
\label{triplet}
\end{table*}

\subsection{Experimental Setup}
In this work, we employ GPT-4o~\cite{hurst2024gpt} as an agent to assist the generation of SurgPub-Video dataset. We train SurgLLaVA-Video on 4 NVIDIA A40 GPUs. 
During the inference, the temperature is set as $0.2$.
We evaluate seven different models on our constructed SurgPub-Video benchmark, including LLaVA-1.5~\cite{liu2023visualllava}, InternVL3~\cite{chen2024internvl}, Qwen-2.5-VL-Instruct~\cite{bai2025qwen2}, GPT-4o~\cite{hurst2024gpt}, Qwen2.5-Max~\cite{bai2025qwen2}, Gemini 2.0 Flash~\cite{team2023gemini}, and our SurgLLaVA-Video.
To further validate the generation ability of our SurgLLaVA-Video model, we follow the experiment setting of SurgVLM~\cite{zeng2025surgvlm} and compare SurgLLaVA-Video with seven models (six benchmarking methods and SurgVLM) on three surgical downstream tasks:
surgical action recognition, surgical skill assessment, and surgical triplet recognition. In particular, we use a multiple-choice question (MCQ) format that requires the model to select the correct answer from the given options and measures the accuracy by counting exact
matches between predictions and ground-truth answers. In each experiment, the best result for each task is highlighted in \textbf{bold}, while the second-best result is indicated with an \underline{underline}. 


\subsection{Benchmark Results}
We evaluate the performance of SurgLLaVA-Video and 
various VLMs in the proposed benchmark. Tab.~\ref{benchmark} presents the
corresponding experimental results. 
SurgLLaVA‑Video achieves consistent and significant improvements across all evaluated organ categories and in overall performance on the benchmark. It reaches 82.84\% overall accuracy, lifting performance by 16.17\% over GPT-4o. 
It obtained more than 10\% improvement in ten of the eleven organs. For cardiac surgery, SurgLLaVA‑Video delivers 84.38\%, outperforming GPT-4o by more than 16\%. Esophageal part obtains 89.69\%, almost 11.56\% higher than the second-best result. 
Moreover, SurgLLaVA-Video achieves these advances despite having substantially fewer parameters with 3B, emphasizing that the effectiveness of incorporating the proposed SurgPub-Video dataset, which
enables the model to develop advanced surgical video understanding capabilities.

\subsection{Downstream Task Results}
\subsubsection{Surgical Action Recognition}
We use SAR-RARP dataset~\cite{psychogyios2023sar} for action recognition, which annotates eight fine-grained robotic actions of prostatectomy. Model performance is measured by classification accuracy, recall, precision, and Jaccard.
Tab.~\ref {rarp&cvs} illustrates the performance of each model on SAR-RARP action recognition dataset.  SurgLLaVA‑Video raises accuracy by more than 22\% over SurgVLM, the strongest competing surgical model, while lifting precision by more than 23\% and the Jaccard index by 14\%. 

\subsubsection{Surgical Skill Assessment}
We benchmark on the critical view of safety (CVS) assessment using the Endoscapes2023 dataset~\cite{mascagni2025endoscapes} from laparoscopic cholecystectomy videos. Each frame is labeled for three CVS criteria: cystic plate exposure, lower gallbladder clearance, and two structures entering the gallbladder. Models predict binary labels for each criterion, with overall accuracy reported for CVS classification. We compute the widely used balance accuracy metrics to assess performance.
As shown in Tab.~\ref{rarp&cvs}, on the Endoscapes2023 CVS task, SurgLLaVA‑Video achieves an average balanced accuracy of 58.88\%, surpassing the strongest baseline, InternVL3, by 5.38\% and outperforming the domain‑adapted SurgVLM by 7.48\%.
A correct CVS assessment requires an understanding of the anatomical structure and surgical operation protocols. The performance improvement of SurgLLaVA-Video indicates its great potential in supporting high-risk clinical decisions. This also proves that the SurgPub-Video dataset can help the model acquire knowledge related to surgeries.

\subsubsection{Surgical Triplet Recognition}
We also evaluate the model performance of triplet recognition on CholecT50~\cite{nwoye2022datatriplet} dataset, which is composed of 50 surgical videos and labeled with the triplet of ``tool-action-target". 
The Mean Average Precision(mAP) and accuracy are used as the main metrics of evaluation. Tab.~\ref{triplet} shows performance of each method on the CholecT50 benchmark. SurgLLaVA‑Video with 3B parameters outperforms SurgVLM by 23.73\% in instrument, 44.65\% in verb, and 26.78\% in triplet, respectively. SurgVLM keeps the top target accuracy at 57.07\%, 9.95\% higher than our model, yet SurgLLaVA‑Video delivers a competitive target mAP of 19.93\% against 21.01\%. 

\subsection{Ablation Study}
Here, we conduct ablation studies to verify the effectiveness of data scale and video length. Because different organs show relatively consistent responses to changes in video length and data scale, whereas each task is more sensitive to these settings, we evaluate model accuracy across five tasks.
\subsubsection{Ablation Study on Video Length}
 Fig.~\ref{ablation}(a) shows MCQ accuracy for the input clip length in 8, 16, 24, and 32 frames. Overall performance peaks at 16 frames with 80.82\%, indicating the best trade‑off between temporal context and redundancy. Instrument recognition and surgery‑planning tasks follow this trend, rising from 8 to 16 frames before tapering. Anatomical structure recognition improves further to 24 frames at 85.08\%, but declines at 32, suggesting that excess frames add noise. Procedure recognition accuracy plateaus after 16 frames, while general knowledge scores drop beyond this length. Overall, 16 frames provide sufficient temporal cues without overwhelming the model, delivering superior accuracy across diverse surgical tasks.
\subsubsection{Ablation Study on Dataset Size}
\begin{figure}[h]
\centering
\includegraphics[width=\linewidth]{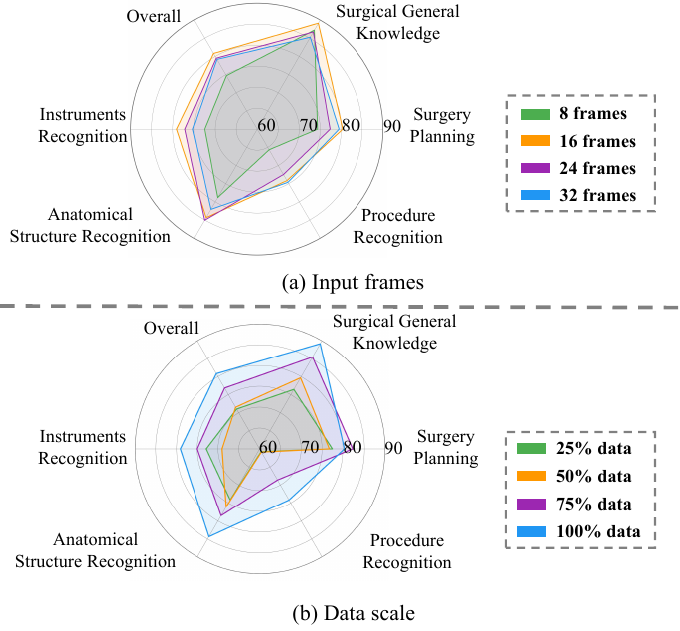}
\caption{Ablation study of SurgLLaVA-Video. (a) Accuracy for 5 tasks and overall score under four input-frame counts. (b) Corresponding accuracy under 4 dataset scales.}
\label{ablation}
\end{figure}

 Fig.~\ref{ablation}(b) reports MCQ accuracy when the model is trained with 25\%, 50\%, 75\%, and 100\% of SurgPub‑Video. Overall accuracy rises almost linearly, reaching 80.82\% with the full dataset. Instrument and Procedure recognition grows by about 7\%–14\% from the smallest to the largest split, showing their dependence on a broad variety of visual exemplars. Anatomical recognition records the steepest improvement, increasing by 10\%, highlighting the importance of diverse anatomical coverage. General‑knowledge accuracy also climbs by 13\%, indicating that richer textual grounding benefits from more clips. Surgery‑planning peaks at 75\% with 82.40\% and then stabilizes, implying diminishing gains once core procedural patterns are captured. These trends confirm that most tasks continue to benefit from larger, well‑balanced data, whereas planning knowledge saturates earlier.

\section{Conclusion}
In this work, we present SurgPub-Video, the first large-scale, video-based surgical dataset specifically designed to enhance VLMs' capabilities for surgical scene understanding. Addressing the existing challenges of limited data availability, semantic scarcity, and unreliable data sources, SurgPub-Video comprises more than 3K peer-reviewed surgical videos and more than 48,000 structured VQA pairs across multiple surgical specialties, enabling video-type input. Building upon this robust dataset, we developed SurgLLaVA-Video, a specialized and efficient surgical VLM architecture. Our model significantly outperformed general-purpose and current surgical-specific models in various benchmarks, achieving state-of-the-art performance despite using only 3 billion parameters. 
Furthermore, we propose a comprehensive multi-task benchmark that systematically evaluates model performance across diverse surgical procedures and clinical tasks.
\bibliography{aaai2026}
\end{document}